\documentclass[aps,prd,preprint,nofootinbib]{revtex4}
\usepackage{amsfonts,amsmath,amssymb}
\usepackage{array}
\usepackage{mathrsfs}
\usepackage{graphicx}
\usepackage{multirow}
\usepackage{epsfig}
\usepackage{graphicx}
\usepackage{subfigure}
\usepackage{booktabs}
\usepackage{array}
\usepackage{tabularx}
\usepackage{slashed}
\usepackage{braket}
\usepackage{bm}
\usepackage{diagbox}
\usepackage{makecell}

\begin{document}
\baselineskip 20pt
\title{Two-loop QCD corrections to $\eta_b \to J/\psi + \gamma$}
\author{\vspace{1cm} Hao Yang$^1$\footnote[1]{yanghao2023@scu.edu.cn}, Lin Zuo$^1$\footnote[2]{2022322020007@stu.scu.edu.cn}, and Bingwei Long$^{1,2}$\footnote[3]{bingwei@scu.edu.cn, corresponding author}\\}

\affiliation{
$^1$College of Physics, Sichuan University, Chengdu, Sichuan 610065, China\\
$^2$Southern Center for Nuclear-Science Theory (SCNT), Institute of Modern Physics, Chinese Academy of Sciences, Huizhou 516000, Guangdong, China\vspace{0.6cm}
}

\begin{abstract}
We present a next-to-leading-order (NLO) analysis of the rare radiative decay $\eta_b \to J/\psi\gamma$, a flavor-changing transition between bottomonium and charmonium, within the framework of non-relativistic QCD (NRQCD). We systematically compute the complete $\mathcal{O}(\alpha_s)$ corrections, which include the one-loop QCD corrections to the QED-initiated amplitudes and the two-loop corrections to the QCD-initiated ones. The branching ratio is enhanced from $2.06^{+2.82}_{-1.32}\times10^{-7}$ at LO to $7.53^{+5.67}_{-1.16}\times10^{-7}$ at NLO, representing an increase by a factor of about 3.65. The theoretical uncertainties caused by renormalization scale and $m_{b/c}$ masses are also discussed. Furthermore, the renormalization scale dependence is reduced at NLO.
\end{abstract}

\maketitle

\section{INTRODUCTION}
As the lowest-lying pseudoscalar bottomonium state, the $\eta_b$ provides an excellent testing ground for the non-relativistic QCD (NRQCD) factorization formalism \cite{Bodwin:1994jh}. Within this framework, nonperturbative hadronization effects are systematically incorporated into long-distance matrix elements (LDMEs), organized according to velocity scaling rules. Given that the relative velocity $v$ for heavy quarks is small, with $v^2 \approx 0.3$ for charmonium and $v^2 \approx 0.1$ for bottomonium, the velocity expansion converges more effectively for bottomonium. This makes bottomonium states particularly suitable for validating the NRQCD factorization approach.

The first evidence for $\eta_b$ is established by BaBar collaboration by observing a peak in the photon energy spectrum of the radiative decay $\Upsilon(3S) \to \eta_b + \gamma$ \cite{BaBar:2008dae}. Previous searches at LEP-II \cite{ALEPH:2002xdz,Abdallah:2006yg}, which rely on multi-hadronic final states, do not yield a significant $\eta_b$ signal, largely attributable to the collider's limited luminosity. In the recent study, the LHCb collaboration searched for the rare decay $\eta_b \to \gamma\gamma$ \cite{LHCb:2025gbn}. No statistically significant signal was observed, leading to an upper limit on the cross-section times branching fraction of $\sigma(pp \to \eta_b X)\times {\rm Br}(\eta_b\to\gamma\gamma) < 765\ \rm pb$ at the 95\% confidence level. Although BaBar has confirmed the existence and mass of $\eta_b$, no measurement of its inclusive or exclusive decay branching ratios has yet been achieved.

The $\eta_b \to \gamma\gamma$ decay, with an estimated branching fraction of $10^{-5}$, represents an ideal channel for $\eta_b$ reconstruction. Nonetheless, its discovery potential is limited by the substantial background in hadron collider from photon pairs produced in multi-$\pi^0$ decays, which constitutes a serious experimental challenge. For the hadronic decay channel $\eta_b \to J/\psi J/\psi$, where both $J/\psi$ mesons subsequently decay to muon pairs, the branching fraction is calculated to be ${\rm Br}(\eta_b \to J/\psi J/\psi) = 8.2 \times 10^{-7}$ with two-loop QCD corrections \cite{Braaten:2000cm, Jia:2006rx, Gong:2008ue, Braguta:2009xu, Sun:2010qx, Zhang:2023mky}. Taking into account the branching fraction for $J/\psi \to \mu\bar{\mu}$ ($\approx$ 6\%) twice, the overall efficiency for the full chain $\eta_b \to J/\psi J/\psi \to 4\mu$ is on the order of $3\times10^{-9}$. This extremely small rate places a high demand on the production yield of $\eta_b$ mesons. 

In this paper, we investigate the decay channel $\eta_b \to J/\psi+\gamma$, whose branching fraction is predicted to be ${\rm Br}(\eta_b \to J/\psi+\gamma) = (1.5\pm0.8)\times10^{-7}$ \cite{Hao:2006nf, Gao:2007fv} at leading order accuracy of NRQCD. This channel features that it requires the reconstruction of only a single $J/\psi$ meson and the purely transverse polarized $J/\psi$ and photon, provides a powerful discriminant for background suppression. Moreover, our two-loop QCD correction reveals a significant enhancement in the branching fraction, due to strong destructive interference between the QED-initiated process (where a $J/\psi$ is produced via a virtual photon in a two-photon diagram) and the QCD-initiated loop diagrams involving gluons. 

Recent technological advances have enabled the calculation of higher-order QCD corrections for quarkonium production and decay, particularly for processes involving multiple quarkonia. For instance, the two-loop corrections to $\Upsilon \to \eta_c(\chi_{cJ})\gamma$ \cite{Zhang:2021ted} and $\Upsilon \to J/\psi\eta_c(\chi_{cJ})$ \cite{Zhang:2022nuf} have been investigated. Notably, the predicted branching fraction for $\Upsilon \to J/\psi\chi_{c1}$ shows good agreement with the Belle measurement \cite{Belle:2014wam}. A substantial two-loop enhancement has also been observed in $\eta_b \to J/\psi J/\psi$ \cite{Zhang:2023mky}, underscoring the critical importance of two-loop effects in such channels. Motivated by these findings, we extend this line of inquiry by computing the $\mathcal{O}(\alpha_s)$ corrections to the radiative decay $\eta_b \to J/\psi\gamma$. This calculation systematically includes the one-loop QCD corrections to the QED-initiated process, as well as the two-loop QCD corrections to the QCD-initiated diagrams, providing a complete next-to-leading-order assessment of this decay mode.

The rest of this paper is organized as follows. In Sect. II, we present the primary formulas employed in the calculation. In Sect. III, the decay widths and their uncertainties are discussed. The last section is reserved for summary and conclusions.

\section{FORMULATION}
The $\eta_b \to J/\psi+\gamma$ decay involves a flavour-changing transition between bottomonium and charmonium states and is therefore highly suppressed in the Standard Model. It can proceed via higher-order diagrams, such as loop-induced diagrams involving gluons, or through two-photon process with one virtual photon converting to $J/\psi$. The principles of parity conservation and Lorentz invariance uniquely determine the tensor structure of the decay amplitude, which must take the following specific form:
\begin{align}
	\mathcal{M}(\lambda_1,\lambda_2) = \mathcal{A}\epsilon_{\mu\nu\alpha\beta}\varepsilon_{J/\psi}^{*\mu}(\lambda_1)\varepsilon_{\gamma}^{*\nu}(\lambda_2)P_1^\alpha P_2^\beta
\end{align}
Here $P_{0,1,2}$ are the momenta of $\eta_b$, $J/\psi$ and $\gamma$, respectively. $\varepsilon_{J/\psi,\gamma}$ are the polarization vector of $J/\psi$ and $\gamma$, $\epsilon_{\mu\nu\alpha\beta}$ is the Levi-Civita tensor. The pseudoscalar nature of the $\eta_b$ imposes a constraint that allows only transverse helicity configurations for the $J/\psi$ and $\gamma$ in its radiative decay, i.e., $(\lambda_1,\lambda_2) = (\pm1,\pm1)$. The reduced amplitude can be expressed as:
\begin{align}
	\mathcal{A} = \frac{1}{2(P_1\cdot P_2)^2} \mathcal{M}^{\mu\nu}\epsilon_{\mu\nu\rho\sigma}P_1^\rho P_2^\sigma,
\end{align}
where $\mathcal{M}(\lambda_1,\lambda_2) = \mathcal{M}_{\mu\nu}\varepsilon_{J/\psi}^{*\mu}(\lambda_1)\varepsilon_{\gamma}^{*\nu}(\lambda_1,\lambda_2)$.

\begin{figure}[htbp!]			
	\centering
	\caption{The typical QED diagrams.}
	\subfigure{\includegraphics[scale=0.8]{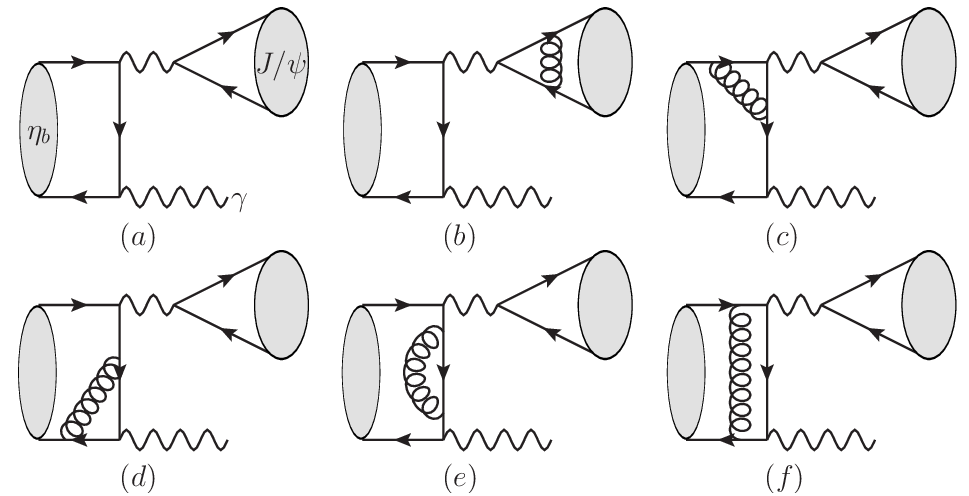}}
	\label{FigQED}
\end{figure}

\begin{figure}[htbp!]			
	\centering
	\caption{The typical QCD diagrams.}
	\subfigure{\includegraphics[scale=0.8]{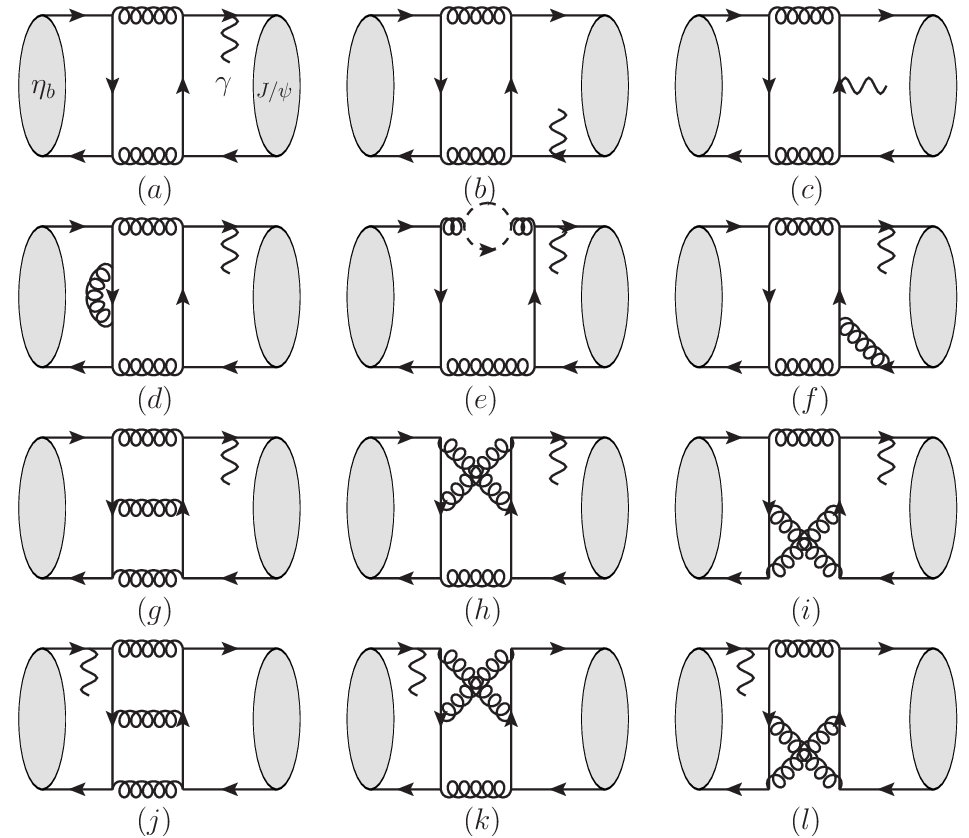}}
	\label{FigQCD}
\end{figure}

According to NRQCD factorization theorem, the production and annihilation of quarkonium can be expressed as partonic processes $b\bar{b} \to c\bar{c} + \gamma$, followed by the projection of $b\bar{b}/c\bar{c}$ into the corresponding quarkonium Fock states. In this process, the color-octet (CO) channel—which requires both the $\eta_b$ and $J/\psi$ to be in a color-octet configuration—is suppressed by $v^8$ and can thus be neglected compared to color-singlet (CS) contributions. The tree-level CS topology, depicted in Fig. \ref{FigQED}-(a), features $\eta_b$ and $J/\psi$ are connected by virtual photon, noted as the QED-initiated diagram. However, due to color conservation and C-parity conservation, the lowest-order QCD diagrams are forbidden at tree level and first contribute at the one-loop level, as exemplified by the representative set shown in Figs. \ref{FigQCD} (a)-(c).

To project the heavy quark pair into a color-singlet state, we employ the standard spin and color projector within the covariant formalism \cite{Bodwin:2002cfe}, for $\eta_b$:
\begin{equation}
	u(p_{b})\bar{v}(p_{\bar{b}})\to \frac{-1}{4\sqrt{2}E_b(E_b+m_b)}(\not\!p_b+m_b)(\not\!P_0+2E_b)\not\! \gamma_{5}(\not\! p_{\bar{b}}-m_b)\otimes(\frac{\bf{1}_{c}}{\sqrt{N_{c}}}),
\end{equation}
while for $J/\psi$:
\begin{equation}
	v(p_{\bar{c}})\bar{u}(p_c)\to \frac{1}{4\sqrt{2}E_c(E_c+m_c)}(\not\! p_{\bar{c}}-m_c)\not\! \epsilon^{*}_{J/\psi}(\not\!P_1+2E_c)(\not\!p_c+m_c)\otimes(\frac{\bf{1}_{c}}{\sqrt{N_{c}}}).
\end{equation}
Here $\bf{1}_{c}$ represents the unit color matrix and $N_c = 3$ is the colors number in QCD. $E_b = \sqrt{P_0^2/4}$ and $E_c = \sqrt{P_1^2/4}$ are the energy of $\eta_b$ and $J/\psi$, $p_Q$ and $p_{\bar{Q}}$ represent the momentum of heavy quark and anti-quark. Since this work is restricted to leading-order accuracy in the relative velocity expansion, we can legitimately neglect the relative momenta between the constituent quarks in both the $\eta_b$ and $J/\psi$. We thus take $p_b = p_{\bar{b}} = P_0/2$ and $p_c = p_{\bar{c}} = P_1/2$.

The total decay width can be formulated as
\begin{align}
	\Gamma[\eta_b\to J/\psi\gamma] = \frac{(m_b^2-m_c^2)^3}{4\pi m_b^3}{\Big|}\mathcal{A}_{\rm QCD}+\mathcal{A}_{\rm QED}{\Big|}^2,
\end{align}
where $\mathcal{A}_{\rm QCD/QED}$ is the reduced amplitude for QCD/QED-initiated diagram, and summation of transverse polarizations for both $J/\psi$ and $\gamma$ is applied. The reduced amplitudes will be calculated perturbatively up to NLO level. Specifically, $\mathcal{A}_{\rm QED}$ and $\mathcal{A}_{\rm QCD}$ are evaluated at the one-loop and two-loop levels, respectively.
\begin{align}
	\mathcal{A}_{\rm QCD} &= - \frac{e_ceg_s^4}{6\pi^2}\sqrt{\frac{m_c}{m_b}}\frac{\psi_{\eta_b}(0)\psi_{J/\psi}(0)}{(m_b^2-m_c^2)^2}\left[ f^{(0)}(r)+\frac{\alpha_s}{\pi}\ f^{(1)}(r) \right], \\
	\mathcal{A}_{\rm QED} &= e_ce_b^2e^3N_c\sqrt{\frac{m_c}{m_b}}\frac{\psi_{\eta_b}(0)\psi_{J/\psi}(0)}{m_c^2(m_b^2-m_c^2)}\left[ 1+\frac{\alpha_s}{\pi}\ h(r) \right],
\end{align}
where $r = \dfrac{m_c^2}{m_b^2}$. Hence, we have
\begin{align}
	\Gamma[\eta_b\to J/\psi\gamma] = \dfrac{64e_c^2\alpha\alpha_s^4}{9}\dfrac{m_c\psi_{\eta_b}^2(0)\psi_{J/\psi}^2(0)}{m_b^4(m_b^2-m_c^2)}{\Big|}f(r)-g(r){\Big|}^2,
\end{align}
where $f(r) = f^{(0)}(r) + \dfrac{\alpha_s}{\pi}f^{(1)}(r)$, $g(r) = g^{(0)}(r) + \dfrac{\alpha_s}{\pi}g^{(1)}(r) = \dfrac{9\pi e_b^2\alpha}{2\alpha_s^2}\dfrac{1-r}{r}{\Big[}1+\dfrac{\alpha_s}{\pi}h(r){\Big]}$. 

We generate the Feynman diagrams and corresponding amplitudes using FeynArts \cite{Hahn:2000kx}. Subsequent algebraic manipulations are performed with FeynCalc \cite{Mertig:1990an}, and a practicable scheme \cite{Korner:1991sx} is applied for handling $\gamma_{5}$ in dimensional regularization. The one- and two-loop integrals are reduced to master integrals (MIs) using FIRE \cite{Smirnov:2014hma}. The one-loop MIs are evaluated analytically via Package-X \cite{Patel:2016fam,Shtabovenko:2016whf}, while the two-loop MIs are computed numerically with AMFlow \cite{Liu:2022chg}. The ultraviolet divergences are canceled by the renormalization constants, while the infrared divergences in exclusive processes cancel each other, ultimately yielding a finite short-distance coefficients (SDCs). To validate our calculation, we first reproduce the leading-order analytical results in Ref. \cite{Hao:2006nf}. Furthermore, some two-loop integrals are cross-checked against FIESTA \cite{Smirnov:2015mct}, with agreement found within the numerical precision of the latter.
 
\section{NUMERICAL RESULTS}
The input parameters are taken as
\begin{align}
	&\alpha = 1/137.065, \hspace{1.7cm} m_c = 1.5\ {\rm GeV}, \hspace{1.5cm} m_b = 4.7\ {\rm GeV},\\ \nonumber	
	&\alpha_s(2m_c) = 0.246,\hspace{1.3cm} \alpha_s(m_b) = 0.211,\hspace{1.3cm} \alpha_s(2m_b) = 0.173\\ \nonumber
	&\psi_{J/\psi}(0) = 0.27\ {\rm GeV}^{3/2},\hspace{0.3cm} \psi_{\Upsilon}(0) = 0.75\ {\rm GeV}^{3/2},\hspace{0.3cm} \psi_{\eta_b}(0) = \psi_{\Upsilon}(0). \nonumber
\end{align}
Here the wave functions at origin for $J/\psi$ and $\Upsilon$ are extracted from their leptonic widths at NLO:
\begin{equation}
	\Gamma(\mathcal{Q}\to e^+e^-)=\frac{4\pi\alpha^2e_Q^2}{m_Q^2}\psi^2_{Q\bar{Q}}(0)\left[1-4C_F\frac{\alpha_s(2m_Q)}{\pi}\right],\ e^{}_Q=\begin{cases}\frac{2}{3},\ {\rm if}\ \mathcal{Q}=J/\psi\\ \frac{1}{3},\ {\rm if}\ \mathcal{Q}=\Upsilon\end{cases},
\end{equation}
with $\Gamma(J/\psi\to e^+e^-)=5.55$ keV, and $\Gamma(\Upsilon\to e^+e^-)=1.34$ keV \cite{ParticleDataGroup:2024cfk}. By virtue of heavy quark spin symmetry in NRQCD at leading order in the relative velocity expansion, we utilize the relation $\psi_{\eta_b}(0) = \psi_{\Upsilon}(0)$. The strong coupling is calculated at $2m_c$ and $2m_b$ using its two-loop formula with $n_f = 4$ and $\Lambda_{\rm QCD} = 297\ \rm MeV$, the intermediate value is chosen at $m_b$.

The total decay width of $\eta_b$ is estimated by its gluonic width at NLO \cite{Bodwin:1994jh}:
\begin{align}
	\Gamma[\eta_b] \approx \Gamma[\eta_b\to gg] = \frac{8\pi\alpha_s^2(2m_b)}{3m_b^2}\psi^2_{\eta_b}(0)\left[1+\frac{\alpha_s(2m_b)}{\pi}(\frac{53}{2}-\frac{31\pi^2}{24}-\frac{8n_f}{9})\right],
\end{align}
i.e., $\Gamma[\eta_b] = 10\ \rm MeV$, consistent with PDG central value $\Gamma[\eta_b] = 10^{+5}_{-4}\ \rm MeV$ \cite{ParticleDataGroup:2024cfk}.

Considering the uncertainties arising from the renormalization scale and the masses of $m_b$ and $m_c$, specifically with $m_b = 4.7\pm0.1\ \rm GeV$, $m_c = 1.5\pm0.1\ \rm GeV$ and the scale $\mu$ varied within the interval $[2m_c, 2m_b]$ (where $m_b$ is taken at its central value), and take into account the two-loop QCD corrections, we find that 
\begin{align}
	{\rm Br}[\eta_b \to J/\psi\gamma] = (7.53^{+5.67}_{-5.16}) \times 10^{-7},
\end{align}
where the LO prediction is $(2.06^{+2.82}_{-1.32}) \times 10^{-7}$.

\begin{table}[ht]
	\caption{The LO and NLO branching fractions for the decay $\eta_b\to J/\psi+\gamma$ from QCD-initiated, QED-initiated and their combined contributions. The renormalization scale $\mu = m_b$, the total decay width of $\eta_b$ is $10\ \rm MeV$.}
	\begin{center}
		\centering
		\begin{tabular}{|m{2.cm}<{\centering}|m{3.cm}<{\centering}|m{3.cm}<{\centering}|m{3.cm}<{\centering}|}
			\toprule
			\hline
			& QCD                 & QED                  & QCD+QED              \\
			\hline
			LO  & $5.18\times10^{-7}$ &  $1.49\times10^{-7}$ &  $2.06\times10^{-7}$ \\
			\hline
			NLO & $20.3\times10^{-7}$ & $0.87\times10^{-7}$  & $7.53\times10^{-7}$  \\
			\hline  		
		\end{tabular}
	\end{center}    
	\label{TabBr}
\end{table}

\begin{figure}[htbp!]			
	\centering
	\subfigure{\includegraphics[scale=0.45]{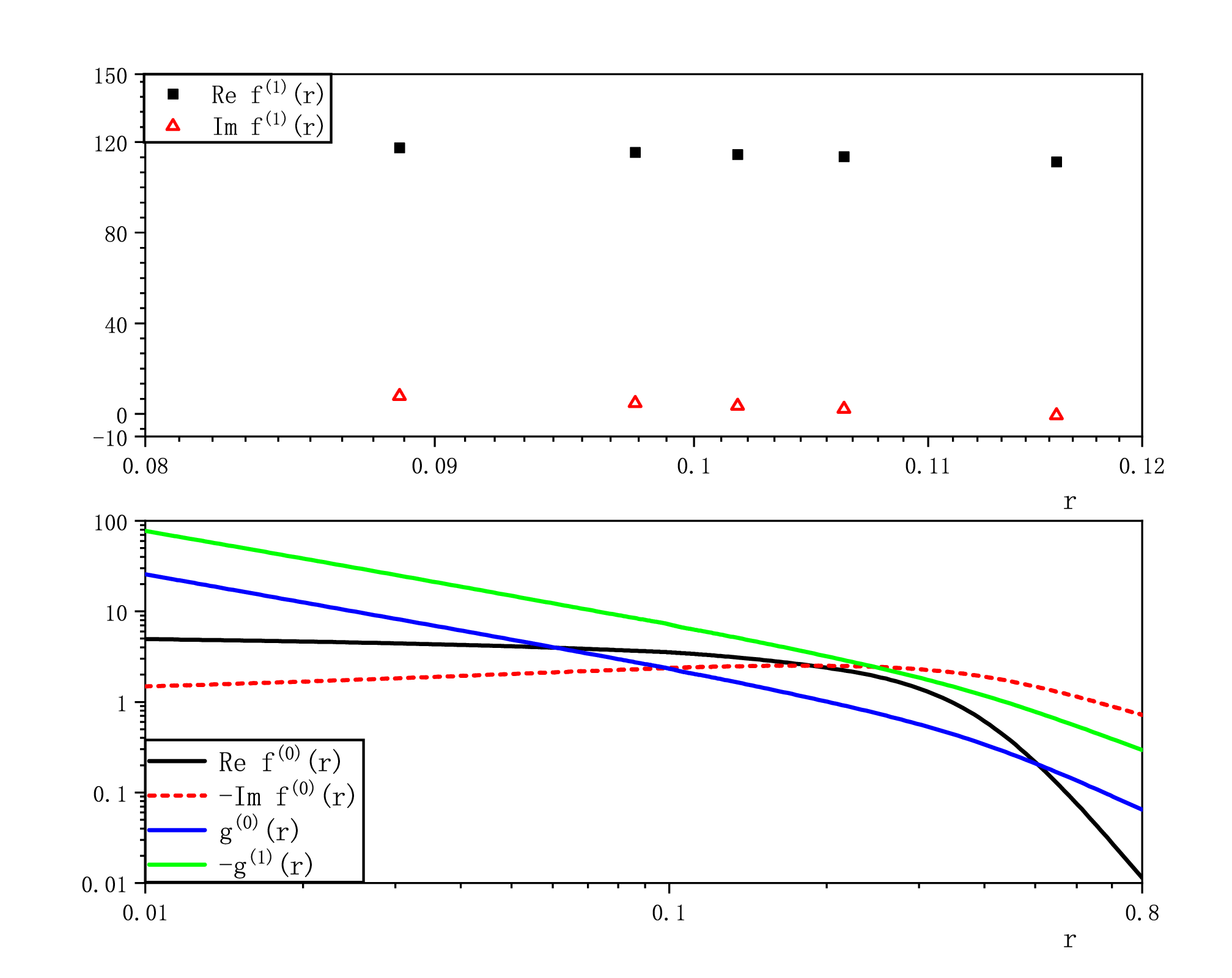}}
	\caption{Real and imaginary part of $f^{(0/1)}(r)$ and $g^{(0/1)}(r)$ from $r=0.01-0.8$, where the two-loop SDC $f^{(1)}(r)$ is given only in several mass ratio, i.e., $r = \frac{(1.5\pm0.1)^2}{(4.7\pm0.1)^2}$.}
	\label{Figfg}
\end{figure} 

The detailed QCD (QED)-initiated results are listed in TABLE \ref{TabBr} and the dimensionless SDCs of $f(r)$ and $g(r)$ are given in  FIG \ref{Figfg}. Numerical prediction based on different $m_c/m_b$ masses settings, e.g., $m_c = 1.45\ \rm GeV$ and $m_b = 4.75\ \rm GeV$, corresponding to $r = 0.093$, can be obtained by interpolating the $f/g$ functions and substituting them into Eq. (8). Although suppressed by a factor of $\alpha/\alpha_s^2$, the QED-initiated contribution enjoys a compensating kinematic enhancement of $m_b^2/m_c^2$ relative to the QCD process. For the LO result, the two $f^{(0)}(r)$ and $g^{(0)}(r)$ functions are comparable in magnitude but are out of phase, leading to destructive interference. Consequently, the squared modulus of their sum, and hence the resultant decay width, is greatly suppressed. Due to the small NLO correction in the QED-initiated channel and the large correction in the QCD-initiated channel, their destructive interference is weakened, e.g., the NLO K-factor for QCD contribution is 3.9 and for total QCD+QED contribution is 3.65. This result contrasts with that of the analogous process $\Upsilon \to \eta_c\gamma$ \cite{Guberina:1980xb,Zhang:2021ted} and the decay $\eta_b \to J/\psi J/\psi$ \cite{Jia:2006rx}, where constructive interference is observed.

\begin{figure}[htbp!]			
	\centering
	\subfigure{\includegraphics[scale=0.5]{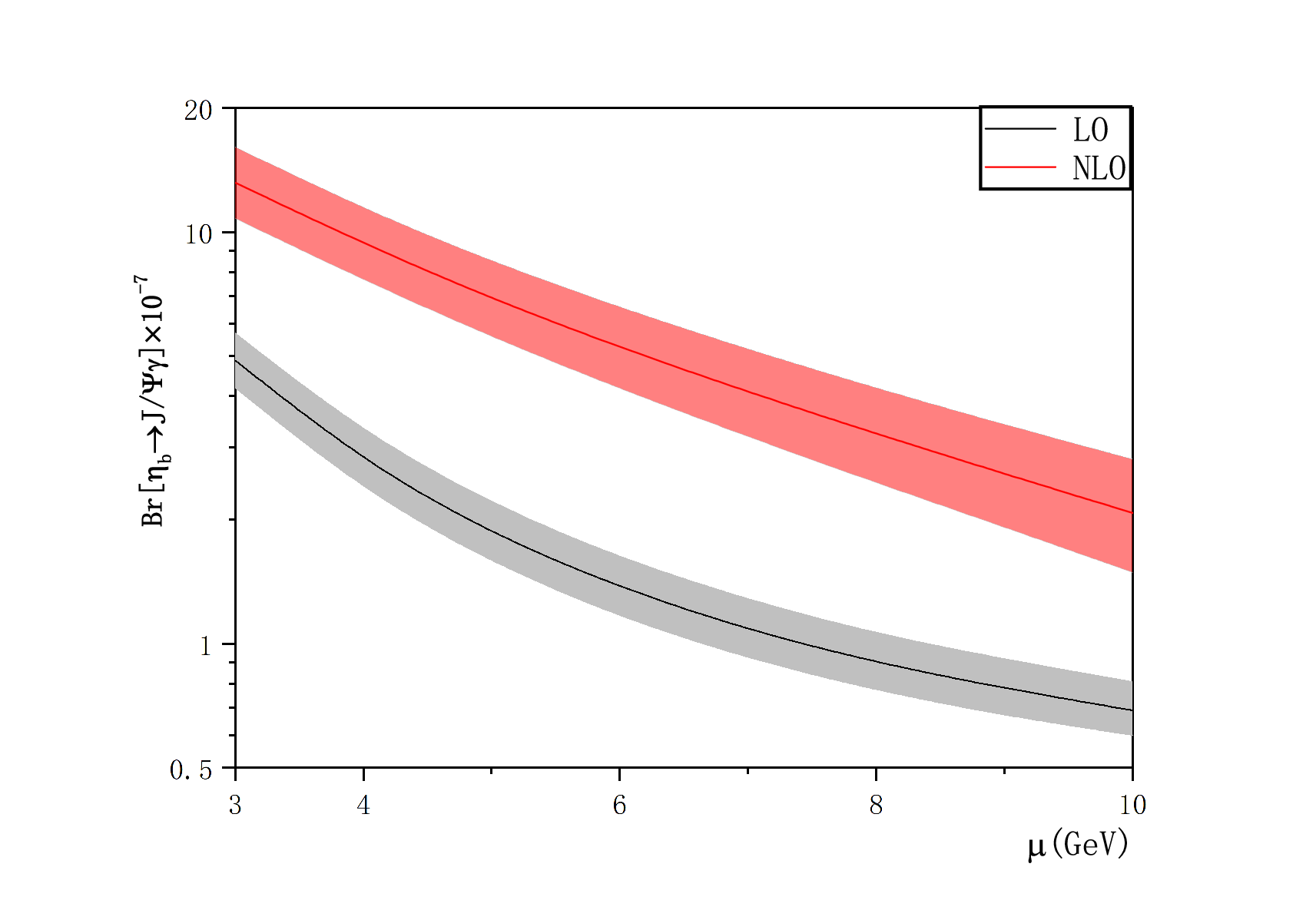}}
	\caption{The branching fraction for $\eta_b\to J/\psi\gamma$ versus renormalization scale $\mu$ with $m_b = 4.7\pm0.1\ \rm GeV$, $m_c = 1.5\pm0.1\ \rm GeV$ and $\Gamma[\eta_b] = 10\ \rm MeV$.}
	\label{FigBr}
\end{figure}

As shown in TABLE \ref{TabBrScale}, the LO QED contribution is scale-independent, and its NLO correction is small. In contrast, the QCD channel varies by a factor of 2–3 across the scale range from $\mu = 2m_c$ to $2m_b$. The destructive interference effect becomes more pronounced at a renormalization scale of $\mu = 2m_b$, leading to $\dfrac{\Gamma_{\rm QCD}}{\Gamma_{\rm QCD+QED}} = 3.19$. The variation of the $m_c$ and $m_b$ masses by $\pm 0.1 \ \rm GeV$ exerts a minor influence on the branching fractions, as detailed in TABLE \ref{TabMassmc} and \ref{TabMassmb}. These variations are accounted for as theoretical uncertainties in Fig.\ref{FigBr}. Notably, the uncertainty arising from the renormalization scale is slightly reduced, evidenced by a flatter slope compared to the LO result.

\begin{table}[ht]
	\caption{The LO (in bracket) and NLO branching fractions for the decay $\eta_b\to J/\psi+\gamma$ versus different renormalization scale from QCD-initiated, QED-initiated and their combined contributions. the total decay width of $\eta_b$ is $10\ \rm MeV$.}
	\begin{center}
		\centering
		\begin{tabular}{|m{2.cm}<{\centering}|m{3.5cm}<{\centering}|m{3.5cm}<{\centering}|m{3.5cm}<{\centering}|}
			\toprule
			\hline
			  $\mu$ & $2m_c$       & $m_b$        & $2m_b$ \\
			\hline
			QCD     & $29.4(9.68)\times10^{-7}$  & $20.3(5.18)\times10^{-7}$  &  $11.6(2.36)\times10^{-7}$\\
			\hline
			QED     & $0.77(1.49)\times10^{-7}$ & $0.87(1.49)\times10^{-7}$ & $0.98(1.49)\times10^{-7}$\\
			\hline
			QCD+QED & $13.2(4.88)\times10^{-7}$  & $7.53(2.06)\times10^{-7}$  & $2.37(0.74)\times10^{-7}$\\
			\hline  		
		\end{tabular}
	\end{center}    
	\label{TabBrScale}
\end{table}

\begin{table}[ht]
	\caption{The LO (in bracket) and NLO branching fractions for the decay $\eta_b\to J/\psi+\gamma$ versus different $m_c$ mass from QCD-initiated, QED-initiated and their combined contributions. Here $m_b = 4.7\ \rm GeV$ and the total decay width of $\eta_b$ is $10\ \rm MeV$.}
	\begin{center}
		\centering
		\begin{tabular}{|m{2.2cm}<{\centering}|m{3.5cm}<{\centering}|m{3.5cm}<{\centering}|m{3.5cm}<{\centering}|}
			\toprule
			\hline
			$m_c$   & 1.4                       & 1.5                       & 1.6                      \\
			\hline
			QCD     & $19.6(4.97)\times10^{-7}$ & $20.3(5.18)\times10^{-7}$ & $21.0(5.36)\times10^{-7}$\\
			\hline
			QED     & $1.09(1.86)\times10^{-7}$ & $0.87(1.49)\times10^{-7}$ & $0.71(1.21)\times10^{-7}$\\
			\hline
			QCD+QED & $5.55(1.68)\times10^{-7}$ & $7.53(2.06)\times10^{-7}$ & $9.32(2.45)\times10^{-7}$\\
			\hline  		
		\end{tabular}
	\end{center}    
	\label{TabMassmc}
\end{table}

\begin{table}[ht]
	\caption{The LO (in bracket) and NLO branching fractions for the decay $\eta_b\to J/\psi+\gamma$ versus different $m_b$ mass from QCD-initiated, QED-initiated and their combined contributions. Here $m_c = 1.5\ \rm GeV$ and the total decay width of $\eta_b$ is $10\ \rm MeV$.}
	\begin{center}
		\centering
		\begin{tabular}{|m{2.2cm}<{\centering}|m{3.5cm}<{\centering}|m{3.5cm}<{\centering}|m{3.5cm}<{\centering}|}
			\toprule
			\hline
			$m_b$   & 4.6                       & 4.7                       & 4.8                      \\
			\hline
			QCD     & $23.7(6.00)\times10^{-7}$ & $20.3(5.18)\times10^{-7}$ & $17.5(4.48)\times10^{-7}$\\
			\hline
			QED     & $0.90(1.55)\times10^{-7}$ & $0.87(1.49)\times10^{-7}$ & $0.84(1.43)\times10^{-7}$\\
			\hline
			QCD+QED & $9.54(2.53)\times10^{-7}$ & $7.53(2.06)\times10^{-7}$ & $5.91(1.68)\times10^{-7}$\\
			\hline  		
		\end{tabular}
	\end{center}    
	\label{TabMassmb}
\end{table}

The total production cross section for $\eta_b$ via gluon-gluon fusion at $\sqrt{s} = 14$ TeV LHC can be estimated by \cite{Maltoni:2004hv}
\begin{align}
	\sigma(pp\to\eta_b+X) = \frac{\pi^2}{64m_b^3}\Gamma[\eta_b\to gg]\frac{d\mathcal{L}_{gg}}{d\mathcal{W}_{gg}}\vert_{\mathcal{W}_{gg}=2m_b},
\end{align}
where the effective gluon-gluon luminosity
\begin{align}
	\frac{d\mathcal{L}_{gg}}{d\mathcal{W}_{gg}} = \int_{\tau}^{1} \frac{\tau}{x}f_g(x,\mu)f_g(\frac{\tau}{x},\mu) dx.
\end{align} 
Here $\tau = \mathcal{W}^2_{gg}/s$, the CTEQ6L1 parton distribution $f_g(x)$ \cite{Pumplin:2002vw} is adopted with $\mu = 2m_b$. The result for the total cross section in $pp$ collision at 14 TeV of center-of-mass energy is $\sigma(pp\to\eta_b+X) = 10\ \rm \mu b$. A detailed prediction \cite{Lansberg:2020ejc,LHCb:2025gbn} in the LHCb acceptance gives $\sigma(pp\to\eta_b+X) = 2.7\ \rm \mu b$.

Based on the preceding discussions regarding the production yield of the $\eta_b$ meson at the LHC, we can now proceed to estimate the phenomenology of the $\eta_b \to J/\psi + \gamma$ decay. This decay mode constitutes an electromagnetic transition and is classified as a rare decay, with an expected branching fraction $7.53\times10^{-7}$. Assuming an integrated luminosity of $\mathcal{L} = 150\ \rm fb^{-1}$ collected during LHC Run 2, and taking into account ${\rm Br}[J/\psi \to \ell \bar{\ell}]$ = 12\% \cite{ParticleDataGroup:2024cfk}, about $1.36\times10^5$ $J/\psi \gamma$ events can be produced. Despite the reduction factors introduced by the finite detection efficiencies for $J/\psi+\gamma$ and the varied kinematic acceptance cuts across experimental setups, the estimated signal yields are projected to remain considerable, paving the way for $\eta_b$ studies at the LHC. Given that the dominant production mechanism of $\eta_b$ at a B Factory is through radiative transitions from Υ(nS) (decay fraction around $5\times10^{-4}$), the total expected yield is merely $10^{5}\ \eta_b$ mesons in approximation. When coupled with the predicted small branching fraction, the number of direct reconstructed signal events becomes prohibitively low. 

\section{SUMMARY AND CONCLUSIONS}
We present a systematic next-to-leading order (NLO) perturbative QCD analysis of the rare radiative decay $\eta_b \to J/\psi\gamma$, a flavor-changing transition between bottomonium and charmonium that is highly suppressed in the Standard Model. Our calculation, performed within the NRQCD factorization framework, incorporates the complete $\mathcal{O}(\alpha_s)$ corrections, which include the one-loop QCD corrections to the QED-initiated process and the two-loop corrections to the QCD-initiated diagrams. The leading-order (LO) result exhibits a significant destructive interference between the comparable QCD and QED amplitudes, leading to a heavily suppressed decay width. Considering the NLO corrections, the QED channel receives only a minor modification, the QCD contribution is enhanced by a large K-factor (approximately 3.9 at $\mu = m_b$), hence the NLO interference behavior is still destructive. The total NLO branching ratio is enhanced to $7.53\times10^{-7}$. Furthermore, the dependency versus renormalization scale is reduced.

The estimated production cross-section for $\eta_b$ via gluon-gluon fusion at the LHC with $\sqrt{s} = 14\ \rm TeV$ is substantial. Combined with our predicted branching fraction and accounting for the $J/\psi \to \ell\bar{\ell}$ decay, the yields is considerable, the future study is feasible. Our results underscore the critical role of higher-order QCD effects in precisely predicting rates for rare quarkonium transitions involving destructive interference and establish $\eta_b \to J/\psi\gamma$ as a compelling channel for future investigation at hadron colliders.

\vspace{1.4cm} {\bf Acknowledgments}
We thank Long-Bin Chen and Zi-Qiang Chen for useful discussion. This work is supported by the National Natural Science Foundation of China (NSFC) under the Grants Nos. 12275185 and 12335002.


\begin{thebibliography}{99}
	\bibitem{Bodwin:1994jh}
	G.~T.~Bodwin, E.~Braaten and G.~P.~Lepage,
	Phys. Rev. D \textbf{51}, 1125-1171 (1995)
	[erratum: Phys. Rev. D \textbf{55}, 5853 (1997)]
	doi:10.1103/PhysRevD.55.5853
	[arXiv:hep-ph/9407339 [hep-ph]].
	
	
	\bibitem{BaBar:2008dae}
	B.~Aubert \textit{et al.} [BaBar],
	Phys. Rev. Lett. \textbf{101}, 071801 (2008)
	[erratum: Phys. Rev. Lett. \textbf{102}, 029901 (2009)]
	doi:10.1103/PhysRevLett.101.071801
	[arXiv:0807.1086 [hep-ex]].
	
	
	\bibitem{ALEPH:2002xdz}
	A.~Heister \textit{et al.} [ALEPH],
	Phys. Lett. B \textbf{530}, 56-66 (2002)
	doi:10.1016/S0370-2693(02)01329-1
	[arXiv:hep-ex/0202011 [hep-ex]].
	
	
	\bibitem{Abdallah:2006yg}
	J.~Abdallah [DELPHI],
	Phys. Lett. B \textbf{634}, 340-346 (2006)
	doi:10.1016/j.physletb.2006.01.058
	[arXiv:hep-ex/0601042 [hep-ex]].
	
	
	\bibitem{LHCb:2025gbn}
	R.~Aaij \textit{et al.} [LHCb],
	[arXiv:2507.14390 [hep-ex]].
	
	
	\bibitem{Braaten:2000cm}
	E.~Braaten, S.~Fleming and A.~K.~Leibovich,
	Phys. Rev. D \textbf{63}, 094006 (2001)
	doi:10.1103/PhysRevD.63.094006
	[arXiv:hep-ph/0008091 [hep-ph]].
	
	
	\bibitem{Jia:2006rx}
	Y.~Jia,
	Phys. Rev. D \textbf{78}, 054003 (2008)
	doi:10.1103/PhysRevD.78.054003
	[arXiv:hep-ph/0611130 [hep-ph]].
	
	
	\bibitem{Gong:2008ue}
	B.~Gong, Y.~Jia and J.~X.~Wang,
	Phys. Lett. B \textbf{670}, 350-355 (2009)
	doi:10.1016/j.physletb.2008.10.063
	[arXiv:0808.1034 [hep-ph]].
	
	
	\bibitem{Braguta:2009xu}
	V.~V.~Braguta and V.~G.~Kartvelishvili,
	Phys. Rev. D \textbf{81}, 014012 (2010)
	doi:10.1103/PhysRevD.81.014012
	[arXiv:0907.2772 [hep-ph]].
	
	
	\bibitem{Sun:2010qx}
	P.~Sun, G.~Hao and C.~F.~Qiao,
	Phys. Lett. B \textbf{702}, 49-54 (2011)
	doi:10.1016/j.physletb.2011.06.060
	[arXiv:1005.5535 [hep-ph]].
	
	
	\bibitem{Zhang:2023mky}
	Y.~D.~Zhang, X.~W.~Bai, F.~Feng, W.~L.~Sang and M.~Z.~Zhou,
	Phys. Rev. D \textbf{108}, no.11, 114030 (2023)
	doi:10.1103/PhysRevD.108.114030
	[arXiv:2310.07453 [hep-ph]].
	
	
	\bibitem{Hao:2006nf}
	G.~Hao, Y.~Jia, C.~F.~Qiao and P.~Sun,
	JHEP \textbf{02}, 057 (2007)
	doi:10.1088/1126-6708/2007/02/057
	[arXiv:hep-ph/0612173 [hep-ph]].
	
	
	\bibitem{Gao:2007fv}
	Y.~J.~Gao, Y.~J.~Zhang and K.~T.~Chao,
	[arXiv:hep-ph/0701009 [hep-ph]].
	
	
	\bibitem{Zhang:2021ted}
	Y.~D.~Zhang, F.~Feng, W.~L.~Sang and H.~F.~Zhang,
	JHEP \textbf{12}, 189 (2021)
	doi:10.1007/JHEP12(2021)189
	[arXiv:2109.15223 [hep-ph]].
	
	
	\bibitem{Zhang:2022nuf}
	Y.~D.~Zhang, W.~L.~Sang and H.~F.~Zhang,
	Phys. Rev. Lett. \textbf{129}, no.11, 112002 (2022)
	doi:10.1103/PhysRevLett.129.112002
	[arXiv:2205.06124 [hep-ph]].
	
	
	\bibitem{Belle:2014wam}
	S.~D.~Yang \textit{et al.} [Belle],
	Phys. Rev. D \textbf{90}, no.11, 112008 (2014)
	doi:10.1103/PhysRevD.90.112008
	[arXiv:1409.7644 [hep-ex]].
	
	
	\bibitem{Bodwin:2002cfe}
	G.~T.~Bodwin and A.~Petrelli,
	Phys. Rev. D \textbf{66}, 094011 (2002)
	[erratum: Phys. Rev. D \textbf{87}, no.3, 039902 (2013)]
	doi:10.1103/PhysRevD.66.094011
	[arXiv:hep-ph/0205210 [hep-ph]].
	
	
	\bibitem{Hahn:2000kx}
	T.~Hahn,
	Comput. Phys. Commun. \textbf{140}, 418-431 (2001)
	doi:10.1016/S0010-4655(01)00290-9
	[arXiv:hep-ph/0012260 [hep-ph]].
	
	
	\bibitem{Mertig:1990an}
	R.~Mertig, M.~Bohm and A.~Denner,
	Comput. Phys. Commun. \textbf{64}, 345-359 (1991)
	doi:10.1016/0010-4655(91)90130-D
	
	
	\bibitem{Korner:1991sx}
	J.~G.~Korner, D.~Kreimer and K.~Schilcher,
	Z. Phys. C \textbf{54}, 503-512 (1992)
	doi:10.1007/BF01559471
	
	
	\bibitem{Smirnov:2014hma}
	A.~V.~Smirnov,
	Comput. Phys. Commun. \textbf{189}, 182-191 (2015)
	doi:10.1016/j.cpc.2014.11.024
	[arXiv:1408.2372 [hep-ph]].
	
	
	\bibitem{Patel:2016fam}
	H.~H.~Patel,
	Comput. Phys. Commun. \textbf{218}, 66-70 (2017)
	doi:10.1016/j.cpc.2017.04.015
	[arXiv:1612.00009 [hep-ph]].
	
	
	\bibitem{Shtabovenko:2016whf}
	V.~Shtabovenko,
	Comput. Phys. Commun. \textbf{218}, 48-65 (2017)
	doi:10.1016/j.cpc.2017.04.014
	[arXiv:1611.06793 [physics.comp-ph]].
	
	
	\bibitem{Liu:2022chg}
	X.~Liu and Y.~Q.~Ma,
	Comput. Phys. Commun. \textbf{283}, 108565 (2023)
	doi:10.1016/j.cpc.2022.108565
	[arXiv:2201.11669 [hep-ph]].
	
	
	\bibitem{Smirnov:2015mct}
	A.~V.~Smirnov,
	Comput. Phys. Commun. \textbf{204}, 189-199 (2016)
	doi:10.1016/j.cpc.2016.03.013
	[arXiv:1511.03614 [hep-ph]].
	

	\bibitem{ParticleDataGroup:2024cfk}
	S.~Navas \textit{et al.} [Particle Data Group],
	Phys. Rev. D \textbf{110}, no.3, 030001 (2024)
	doi:10.1103/PhysRevD.110.030001
	
	
	\bibitem{Guberina:1980xb}
	B.~Guberina and J.~H.~Kuhn,
	Lett. Nuovo Cim. \textbf{32}, 295 (1981)
	doi:10.1007/BF02745123
	
	
	\bibitem{Maltoni:2004hv}
	F.~Maltoni and A.~D.~Polosa,
	Phys. Rev. D \textbf{70}, 054014 (2004)
	doi:10.1103/PhysRevD.70.054014
	[arXiv:hep-ph/0405082 [hep-ph]].
	
	
	\bibitem{Pumplin:2002vw}
	J.~Pumplin, D.~R.~Stump, J.~Huston, H.~L.~Lai, P.~M.~Nadolsky and W.~K.~Tung,
	JHEP \textbf{07}, 012 (2002)
	doi:10.1088/1126-6708/2002/07/012
	[arXiv:hep-ph/0201195 [hep-ph]].
	
	
	\bibitem{Lansberg:2020ejc}
	J.~P.~Lansberg and M.~A.~Ozcelik,
	Eur. Phys. J. C \textbf{81}, no.6, 497 (2021)
	doi:10.1140/epjc/s10052-021-09258-7
	[arXiv:2012.00702 [hep-ph]].
	


\end{thebibliography}
\end{document}